\documentclass[11pt,twoside]{article}
\usepackage{CAGN2019}
\usepackage{graphicx}

\usepackage[T1]{fontenc} 

\usepackage{latexsym}
\usepackage{verbatim}

\usepackage{ifpdf}  
\ifpdf  
      \DeclareGraphicsExtensions{.pdf,.png,.jpg}  
\else  
      \DeclareGraphicsExtensions{.eps}  
\fi 

\begin{document}
\newcommand\ion[2]{#1$\;${\scshape{#2}}}

\vskip 1.0cm
\markboth{K. Z.~Arellano-C\'ordova et al.}{Abundance gradients in the Milky Way}
\pagestyle{myheadings}
%
%
\vspace*{0.5cm}
\parindent 0pt{Poster}


\vspace*{0.5cm}
\title{The radial abundance gradients of O, Ne, S and Cl of the Milky Way}

\author{K. Z.~Arellano-C\'ordova$^{1,2,3}$, C.~Esteban$^{1,2}$ and J.~Garc\'ia-Rojas$^{1,2}$}

\affil{$^1$Instituto de Astrof\'{\i}sica de Canarias, E-38200 La Laguna, Tenerife, Spain \\
$^2$Departamento de Astrof\'{\i}sica, Universidad de La Laguna, E-38206 La Laguna, Tenerife, Spain\\
$^3$ Instituto Nacional de Astrof\'isica, \'Optica y Electr\'onica (INAOE),
    Apdo. Postal 51 y 216, Puebla, Mexico\\
}

\begin{abstract}
We present preliminary results of the O, Ne, S and Cl abundance gradients of the Milky Way. We analyze in a homogenous way the physical conditions and chemical abundances of a sample of 35 \ion{H}{ii} regions with deep spectra observed mainly with the GTC and VLT telescopes. The sample covers a range in Galactocentric distances from 5 to 17 kpc. We reanalyze the O abundance gradient, obtaining a similar complex shape than previously reported in the literature. We calculate the Cl abundance gradient including a significantly larger number of objects than in previous works. Our results show a gradient for Cl/H of $-$0.034 dex kpc$^{-1}$ and a dispersion around the gradient of 0.14 dex. We obtain values for the Ne/H and S/H abundance gradients with slopes of $-$0.038 dex kpc$^{-1}$ and $-$0.046 dex kpc$^{-1}$, respectively, with dispersions around the gradient higher than 0.25 dex. We also report the values for the slopes of the Cl/O, Ne/O and S/O abundance ratio gradients, which show dispersions around the gradient up to 0.4 dex. Such high dispersion although may be interpreted as indications of chemical inhomogeneities in the Galactic ISM, it might be also an artifact produced by the selected ionization correction factor. 


\end{abstract}

\section{Introduction}
The Milky Way is an excellent laboratory to study chemical abundances of \ion{H}{ii} regions as a function of their Galactocentric distances since it helps to constrain chemical evolution models. Deep spectra from Galactic \ion{H}{ii} regions are necessary to have a better estimate of the electron temperature and to derive precise ionic and total abundances. Such high quality observations allow to study in detail the temperature structure of the nebula, the ionization corrector factors (ICF), which are used to account for the contribution of unseen ions in the calculation of the total abundance, and the abundance gradients of most of the observable elements. It is important the use of ICF to calculate the total abundances of elements such as C, N, Ne, S, Cl and Ar. 
Several studies have proposed ICF based on the ionization potential of given ions, photoionization models or observational data (e.g. \citealt{Peimbert:1969,Izotov:2006, Delgado-Inglada:2014, Esteban:2015, Esteban:2018}). 

On the other hand, the radial abundance gradients in the Milky Way of elements as Cl, O and N have been studied in detail by \citet{Esteban:2015, Esteban:2017} and \citet{Esteban:2018}. In the case of Cl, \citet{Esteban:2015} reported values for the gradients of Cl/H and Cl/O for a sample of nine objects without using ICF. These authors found similar slopes between Cl/H and O/H of $-$0.043 dex kpc$^{-1}$ and a flat slope for the Cl/O abundance gradient. Recently, \citet{Esteban:2018} reported a new determination for the N/H abundance gradient, where for objects with very low ionization degree they assume the approximation N/H $\approx$ N$^{+}$/H$^{+}$ (N$^{++}$ contribution is expected to be negligible due to the low ionization degree). \citet{Esteban:2018} compared three different schemes of ICF concluding that there are not large differences or trends between the objects where ICF was used and those ones without using ICF.  Note that the ICF reported in the literature are based mainly on either photoionization models of extragalactic \ion{H}{ii} regions or planetary nebulae. Therefore, it is necessary to analyze the applicability of ICF in Galactic \ion{H}{ii} regions to have a better estimate of the total abundance in these objects. In this work, we present preliminary results on the Ne, S and Cl abundance gradients and a reanalysis of the O abundance gradient in the Milky Way by using a sample of deep spectra of \ion{H}{ii} regions.

\section{Sample}
We gathered a sample of  35 \ion{H}{ii} regions observed mainly with the Gran Telescopio Canarias (GTC) in La Palma, Spain and the Very Large Telescope (VLT) in Chile from \citet{ Garcia-Rojas:2014, Esteban:2015, Esteban:2017, Fernandez-Martin:2017, Esteban:2018}. The Galactocentric distances for our sample go from 5 to 17 kpc. 

\section{Physical conditions and chemical abundances}
 The physical conditions and chemical abundances were computed in a homogenous way using the same atomic data set than \citet{Esteban:2018}, and  by using the Python package PyNeb (\citealt{Luridiana:2015}). We adopt a two-zone ionization structure characterized by $T_{\rm e}$[\ion{N}{ii}] for O$^{+}$, S$^{+}$ and Cl$^{+}$ and by $T_{\rm e}$[\ion{O}{iii}] for O$^{++}$, Ne$^{++}$, S$^{++}$, Cl$^{++}$ and Cl$^{3+}$ ionic abundances. We also use the temperature relation proposed by \citet[in their equation 3]{Esteban:2009}  to estimate $T_{\rm e}$[\ion{O}{iii}] in those objects where this temperature was not available.  

The total abundances of O, Cl, Ne, and S were calculated as follows: for O/H, we add the contribution of both ions, O/H = O$^+$/H$^+$ + O$^{++}$/H$^+$. For the other elements, it was necessary the use of ICF. For Cl, we use the ICF of \citet{Izotov:2006}, which depends on metallicity. Note that in the sample compiled by \citet{Esteban:2015}, the total Cl abundance was determined without the use of ICF. In the case of Ne, we use the ICF reported by \citet{Peimbert:1969}, which is based on the similar ionization potentials of Ne$^{++}$ and O$^{++}$. For S, we use the ICF(S$^+$ + S$^{++}$) reported by \citet{Izotov:2006}, which also depends on metallicity. 

\section{Preliminary results}

We reanalize the O abundance gradient in the Milky Way as a function of its Galactocentric distance for the whole sample of  \ion{H}{ii} regions. We obtain a similar O/H gradient than that previously reported by \citet{Esteban:2018}, with a dispersion of 0.08 dex at a given Galactocentric distance. 

\begin{figure}[h]  
\begin{center}
\includegraphics[height=3.8cm]{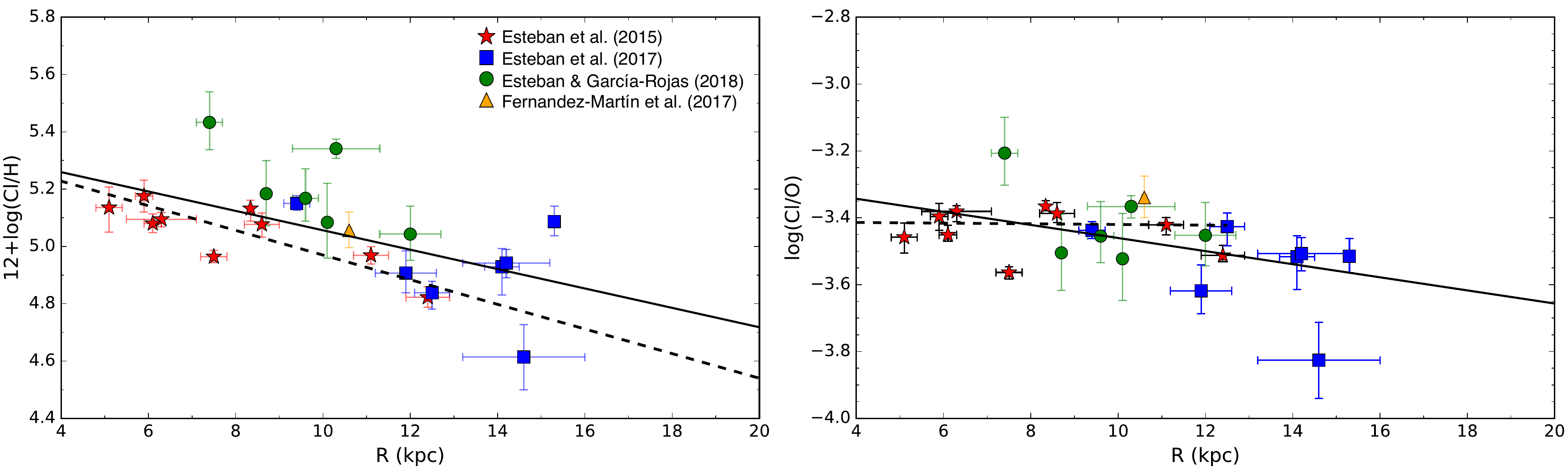}
\caption{Left: Cl abundances in \ion{H}{ii} regions of the Milky Way as a function of their Galactocentric distances. Right: The Cl/O abundances as a function of their Galactocentric distances. Note that the Cl abundances for the sample of \ion{H}{ii} regions of \citet{Esteban:2015} were computed without the use of ICF. The solid line represents our fit to all the objects and the dashed line the radial abundances of Cl/H and Cl/O reported by \citet{Esteban:2015}.}
\label{fig1}
\end{center}
\end{figure}

Figure \ref{fig1} shows the Cl/H and Cl/O abundance gradients from the \ion{H}{ii} regions in the Milky Way as a function of their Galactocentric distances. The different symbols indicate the references in the literature where the samples were compiled. We used the least-squares method to fit all the objects in Figure \ref{fig1}. Our results show that the slope of the Cl/H gradient is slightly flattened in comparison to the one reported by \cite{Esteban:2015} with a dispersion of 0.14 dex. We report a slope of $-$0.020 dex for the Cl/O gradient and a dispersion around the gradient of 0.10 dex. However, \citet{Esteban:2015} found a flat Cl/O gradient for objects located between 5 and 12 kpc, the expected behaviour because the bulk of both elements -- O and Cl -- are thought to be produced by massive stars. We will explore the effect of the selected ICF onto this apparent odd behaviour.  The Ne and S abundance gradients show a large dispersion around the gradient of 0.25 dex and 0.45 dex, respectively, which also might be due to the selected ICF. We plan to study in detail the different ICF proposed in the literature for these elements, and hence, to improve the chemical abundance determinations and have a better estimate of the gradient.  We also calculated the Ne/O and S/O abundance gradients, which shows slopes of $-$0.008 dex kpc$^{-1}$ and $-$0.016 dex kpc$^{-1}$ and dispersions of 0.22 dex and 0.40 dex, respectively. It is curious that these elements also show somewhat negative slopes as that of the Cl/O ratio. The reasons of this fact will be further investigated.

\section{Summary and work in progress}
We present preliminary results of the Ne, S and Cl abundance gradients of the Milky Way. 
We recalculate the physical conditions and obtain new determinations of the chemical abundances of these elements for a sample of Galactic \ion{H}{ii} regions with available deep, high-quality spectra. We obtain similar values for the abundance gradients than those reported in the literature. Our determinations for the total abundance gradients show a large dispersion around the gradient up to 0.45 dex (ICF problems?). For upcoming research, we plan to analyze the argon and carbon abundances and explore in detail the ICF, mainly those proposed by Medina-Amayo (in prep) for \ion{H}{ii} regions based on photoionization models and the ICF for Cl based on observational data by Dom\'inguez-Guzm\'an (in prep). 

\acknowledgments KZA-C acknowledges support from Mexican CONACYT grant 364239. CEL and JRG acknowledge support from the project AYA2015-65205-P. JGR acknowledges support from an Advanced Fellowship from the Severo Ochoa excellence program (SEV-2015-0548).

\bibliographystyle{aaabib}
\bibliography{Arellano-Cordova}

\end{document}